\documentclass[12pt]{article}
\usepackage{amssymb}

\usepackage[top=2.75cm, bottom=2.75cm, left=2.75cm, right=2.75cm]{geometry} 

\newcommand{\be}{\begin{equation}}
\newcommand{\ee}{\end{equation}}
\newcommand{\beqs}{\begin{eqnarray}}
\newcommand{\eeqs}{\end{eqnarray}}
\newcommand{\Lag}{\mathcal{L}}

\newcommand{\al}{\alpha}

\newcommand{\ie}{{\it i.e.},~}

\def\({\left(}
\def\){\right)}

\def\ni{\noindent}

\def\mxth{\mathsurround=0pt }
\def\xversim#1#2{\lower2.pt\vbox{\baselineskip0pt \lineskip-.5pt
x  \ialign{$\mxth#1\hfil##\hfil$\crcr#2\crcr\sim\crcr}}}

\newcommand{\pa}{\partial}

\newcommand{\q}{\theta}

\def\be{\begin{equation}}
\def\ee{\end{equation}}
\def\bea{\begin{eqnarray}}
\def\eea{\end{eqnarray}}

\DeclareSymbolFont{AMSb}{U}{msb}{m}{n}
\DeclareMathSymbol{\fieldC}{\mathalpha}{AMSb}{"43}
\DeclareMathSymbol{\fieldR}{\mathalpha}{AMSb}{"52}
\DeclareMathSymbol{\fieldZ}{\mathalpha}{AMSb}{"5A}

\begin{document} 

\setcounter{page}{0}
\thispagestyle{empty}

\begin{flushright} \small
YITP--SB--05--39 \\ HUTP-05/A055 \\ ITP--UU--05/56 \\ SPIN--05/36 \\
\end{flushright}
\smallskip

\begin{center} \LARGE
 Hypermultiplets and Topological Strings
 \\[6mm] \normalsize
 Martin Ro\v{c}ek$^1$, Cumrun Vafa$^2$ and Stefan Vandoren$^3$ \\[3mm]
 {\small\slshape
 $^1$C.N. Yang Institute for Theoretical Physics\\
 State University of New York at Stony Brook, NY 11790, USA\\[3mm]
 $^2$Jefferson Physical Laboratory, Harvard University\\
 Cambridge, MA 02138, USA \\[3mm]
 $^3$Institute for Theoretical Physics \emph{and} Spinoza Institute \\
 Utrecht University, 3508 TD Utrecht, The Netherlands}
\end{center}
\vspace{10mm}

\centerline{\bfseries Abstract} \medskip

\ni The c-map relates classical hypermultiplet moduli spaces in 
compactifications of type II strings on a Calabi-Yau threefold
to vector multiplet moduli spaces via a 
further compactification on a circle. We give an off-shell description 
of the c-map in $N=2$ superspace. The superspace Lagrangian for the 
hypermultiplets is a single function directly related to 
the prepotential of special geometry, and can therefore be computed using
topological string theory. Similarly, a class of higher derivative terms
for hypermultiplets can be computed from the higher genus topological 
string amplitudes. Our results provide a framework for studying
quantum corrections to the hypermultiplet moduli space, as well as for 
understanding the black hole wave-function as a function of the hypermultiplet
moduli. 

\eject

\small
\vspace{1cm}
\tableofcontents{}
\vspace{1cm}
\bigskip\hrule
\normalsize

\section{Introduction}
\setcounter{equation}{0}

String theories with eight supercharges have been studied intensely 
over recent times, and still provide an excellent laboratory to study 
various phenomena that are important in any theory of quantum gravity.
Perhaps the most studied examples are type II superstrings compactified 
on a Calabi-Yau threefold, and the connection to topological strings 
\cite{Bershadsky:1993cx,Antoniadis:1993ze}. 
In this context, it was recently conjectured that the partition functions of 
BPS black holes and topological strings are related in a simple way 
\cite{Ooguri:2004zv}. This conjecture was supported by the supergravity
calculations done in \cite{LopesCardoso:1998wt}, where subleading corrections 
to the BPS entropy coming from certain higher derivative 
terms in the supergravity effective action were determined.

While most aspects of black hole physics are related to the vector multiplet
moduli space, there is also the hypermultiplet moduli space which is
much less well understood. This is because 
the string coupling constant $g_s$ sits in a hypermultiplet and therefore, 
this sector is subject to quantum corrections in $g_s$, both perturbatively
and non-perturbatively through five-brane and membrane instantons \cite{BBS}. 
Not much is known about these corrections, 
and this is mainly due to the complicated nature of the 
quaternion-K\"ahler (QK) geometry underlying the hypermultiplet moduli space.
Some results can be obtained when restricting to the sector of the 
universal hypermultiplet only 
\cite{Antoniadis:2003sw,Anguelova:2004sj,Davidse:2004gg,Davidse:2005ef}, or 
near a conifold singularity \cite{Greene:1996dh,Ooguri:1996me}.

At the classical level, however, the hypermultiplet moduli space is 
well known to be related to the special geometry of the vector multiplet
sector through the c-map \cite{CFG,FS}. At the supergravity level, the 
c-map arises upon compactifying the four-dimensional supergravity action on
a circle, after which vector multiplets can be dualized into hypermultiplets.
This means that the quaternionic geometry for the hypermultiplets is 
completely 
determined by a single function, namely the holomorphic prepotential of 
special geometry, or equivalently, the genus zero topological string 
amplitude. As will become clear in this paper, by using the conformal tensor 
calculus developed in \cite{DWKV,DWRV}, 
the c-map has a natural and simple off-shell 
description in $N=2$ projective superspace \cite{PSS1,PSS2}.
We show that the superspace Lagrangian corresponding to the 
quaternionic geometry simply amounts to integrating
the topological string amplitude over projective superspace. 
Moreover, we propose a relation between the higher genus topological 
string amplitudes and certain higher derivative terms in the effective 
action for the hypermultiplets. Similar terms were written 
down in components in \cite{Antoniadis:1993ze}.
Superspace effective actions, in relation to the c-map and topological 
strings, were also studied in \cite{BS,Berko}.

Initially, our motivation came from understanding stringy corrections 
to the hypermultiplet moduli space in terms of topological strings. 
Using superspace techniques, both perturbative and nonpertubative corrections
can be encoded by a single function that determines the entire
Lagrangian and quaternionic geometry \cite{DWRV}. This is illustrated in 
\cite{Anguelova:2004sj,RSV} for the perturbative corrections. 
It remains to be understood if these corrections, and in particular the 
membrane instanton corrections, can also be calculated using topological 
strings. More recently, it was shown in \cite{OVV} that there is also
a connection to black hole physics. There, it was argued that the   
Hartle-Hawking wave function for BPS black holes should be understood as 
a function of all moduli that appear after a further compactification of 
the four-dimensional theory on a circle. These moduli naturally sit in 
hypermultiplets and therefore, the wave function is a function defined 
on the QK manifold, precisely as introduced by the c-map. See also 
\cite{Pioline}. We comment on these issues further in Section 5.

This paper is organized as follows. In Section 2, we review the c-map
and describe the associated quaternionic geometries. In Section 3, we 
state our main result, namely the superspace Lagrangian 
for the c-map, which we prove in Section 4. We use
various important geometrical features that are related to QK
manifolds, namely their twistor spaces, and their hyperk\"ahler cones. In Section 5, we 
summarize our results and suggest possible further connections with black holes
and topological strings. Finally, in Section 6, we propose a class of 
higher derivative terms that should relate to higher genus topological string 
amplitudes.

\section{The c-map}\label{cmap}
\setcounter{equation}{0}

In this section, we introduce our notation and review the
c-map originally constructed in \cite{CFG,FS}.

Low-energy effective actions for type II strings on Calabi-Yau (CY) threefolds
contain both vector multiplets and hypermultiplets.  The $N=2$ supergravity
couplings require the scalars of the vector multiplets to parametrize a
special K\"ahler manifold \cite{deWit:1984pk}, whereas the hypermultiplet 
scalars parametrize a Quaternion-K\"ahler manifold \cite{Bagger:1983tt}.

The projective (or rigid) special K\"ahler manifold has real dimension
$2(n+1)$ and is characterized by a holomorphic prepotential $F(X^I)$, which
is homogeneous of degree two ($I=1,\cdots,n+1$). In type IIA (IIB)
compactifications on a CY, we have $n=h_{1,1}\, (h_{1,2})$, respectively.

The K\"ahler potential and metric of the rigid special geometry are given
by\footnote{We use the modern conventions for the prepotential. In
the original references \cite{FS,DWVVP},
different conventions were used: $K=\frac{1}{4}\Big(X^I{\bar F}_I
+{\bar X}^IF_I\Big)$ and $N_{IJ}=\frac{1}{4}(F_{IJ}+{\bar F}_{IJ})$.
It is straightforward to switch between these conventions.}
\begin{equation}\label{K-pot}
K=i({\bar X}^IF_I-X^I{\bar F}_I)\ ,\qquad
N_{IJ}=i(F_{IJ}-{\bar F}_{IJ})\ ,
\end{equation}
where $F_I$ is the first derivative of $F$, etc. In terms of the
periods of the Calabi-Yau manifold, for type IIB theory, we may identify 
\begin{equation}\label{Om-periods}
X^I=\int_{A_I}\,\Omega \ ,\qquad F_I =\int_{B^I}\,\Omega\ ,
\end{equation}
where $A_I$ and $B^I$ are a real basis of three-cycles, $I=0,...,h_{1,2}$,
and $\Omega$ is the holomorphic three-form. The Riemann
bilinear identity implies that the K\"ahler potential (\ref{K-pot}) is
\begin{equation}\label{K-CY}
K=-i\int_{CY}\,{\Omega} \wedge {\bar \Omega}\ .
\end{equation}

The (local) special K\"ahler geometry is
then of real dimension $2n$, with complex inhomogeneous coordinates
\begin{equation}\label{proj-coord}
Z^I=\frac{X^I}{X^1} = \{1,Z^A\}\ ,
\end{equation}
where $A$ runs over $n$ values. Its K\"ahler potential is given
by
\begin{equation}\label{local-K}
{\cal K}= {\rm ln} (Z^IN_{IJ}{\bar Z}^I)\ .
\end{equation}
We further introduce the matrices \cite{deWit:1984pk}
\begin{equation}\label{curlyN}
{\cal N}_{IJ}=-i{\bar F}_{IJ}- \frac{(NX)_I(NX)_J}{(XNX)}\ ,
\end{equation}
where $(NX)_I\equiv N_{IJ}X^J$, etc.  These matrices determine the gauge
kinetic terms in the vector multiplet action and completely specify the 
couplings of vector multiplets to $N=2$ supergravity in four spacetime 
dimensions. 

The classical c-map is found by compactifying from four to three 
dimensions on a circle $S^1$. Each gauge 
field in four dimensions yields a pair of massless scalars in three dimensions:
one comes from the component of the four dimensional gauge field along the circle, 
and the other from dualizing the remaining three-dimensional 
gauge field into a scalar.

Doing so, one maps a vector multiplet into a hypermultiplet, which
we schematically denote by
\begin{equation}
{\rm c-map}:\quad (Z^A, A_\mu^I) \rightarrow (Z^A, A^I, A_{\hat \mu}^I)
\rightarrow (Z^A,A^I,B_I)\ .
\end{equation}
(Alternatively, we note that in three-dimensions, a vector multiplet is equivalent to a tensor
multiplet. Then the c-map can be regarded as taking a vector multiplet from four to three
dimensions and reinterpreting it as a tensor multiplet when returning to four dimensions.
The tensor multiplet can then be dualized into a hypermultiplet in four dimensions.)
In addition to the scalars arising from the gauge fields, two more scalars 
$\phi$ and $\sigma$ come from the metric tensor, so we find a
total of $4(n+1)$ scalars.

After the c-map, we obtain hypermultiplets whose scalars parametrize a
target space that is a Quaternion-K\"ahler manifold of dimension $4(n+1)$.
These spaces were described in \cite{FS},
and further analyzed in \cite{DWVVP}. We use the notation of the latter
reference (replacing
$\phi$ by ${\rm e}^\phi$). The QK metric can then be written as
\begin{eqnarray}\label{QK-metric}
{\rm d}s^2&=&{\rm d}\phi^2 - {\rm e}^{-\phi}({\cal N}+{\bar {\cal N}})_{IJ}
W^I{\bar W}^J+{\rm e}^{-2\phi}\Big( {\rm d}\sigma 
-\frac{1}{2}(A^I{\rm d}B_I-B_I{\rm d}A^I)\Big)^2
\nonumber\\
&&-4 {\cal K}_{A\bar B}\,{\rm d}Z^A {\rm d}{\bar Z}^{\bar B}\ .
\end{eqnarray}
The metric is only positive definite in the domain where
$(ZN{\bar Z})$ is positive and hence ${\cal K}_{A{\bar B}}$ is negative
definite. One can then show that
${\cal N}+{\bar {\cal N}}$ is negative definite \cite{CKVPDFDWG}.  The
one-forms $W^I$ are defined by
\begin{equation}
W^I=({\cal N}+{\bar {\cal N}})^{-1\,IJ}\Big(2{\bar {\cal N}}_{JK}
{\rm d}A^K-i{\rm d}B_J\Big)\ .
\end{equation}

Although we have constructed the QK manifold from a supergravity action
dimensionally reduced to three dimensions, one can write down four-dimensional
supergravity lagrangians coupled to hypermultiplets that parametrize
the same QK manifold as in (\ref{QK-metric}). These are precisely the ones
that appear in CY compactifications \cite{CFG,FS}. The underlying mechanism
is that T-duality
\begin{equation}
{\rm IIA}/\Bigl({\rm CY}\times S^1_R\Bigr) \simeq {\rm IIB}/\Bigl({\rm CY}
\times S^1_{1/R}\Bigr)
\end{equation}
relates type IIA and IIB string theory compactified on the {\it same}
CY manifold.

Finally, the scalar field $\phi$ in (\ref{QK-metric}) is identified with the
dilaton and arises from the purely gravitational sector after the c-map.
It belongs to the universal hypermultiplet. In our conventions, the relation
with the string coupling constant is given by
\begin{equation}\label{gstring}
g_{s}\equiv {\rm e}^{-\frac12\phi_\infty}\ ,
\end{equation}
where $\phi_\infty$ is the value of the dilaton at infinity.

\section{Superspace description and Legendre transform}
\label{Superspace}
\setcounter{equation}{0}

In this section we give the superspace description of the Lagrangian
corresponding to the QK metric (\ref{QK-metric}). Off-shell descriptions
of matter couplings in $N=2$ supergravity can be conveniently formulated
using the superconformal tensor calculus. For hypermultiplets this tensor
calculus was developed in \cite{DWKV,DWRV}. The geometry of the scalar
manifolds is again projective, as for special K\"ahler manifolds. The
compensators restoring the dilatations and $SU(2)_R$ inside the conformal
group form an entire hypermultiplet. Adding the compensator to the original
hypermultiplets that parametrize the $4(n+1)$-dimensional QK space, one
obtains a parent space of dimension
$4(n+2)$. This space is actually hyperk\"ahler, admits a homothety
and a $SU(2)$ isometry group that rotates the three complex structures.
In the mathematics literature, this space is called the Swann space
\cite{Swann}. In the physics literature, we have used the name
{\it hyperk\"ahler cone} (HKC) \cite{DWRV}.

\subsection{Projective superspace}

The lagrangian of an HKC corresponds to an $N=2$ conformally invariant
supersymmetric sigma model. Off shell, actions for such models can be
conveniently written in terms of an integral in projective superspace
\cite{PSS1,PSS2},
\begin{equation}
S= {\rm Im} \int\,{\rm d}^4 x \; {\rm d}^2\theta {\rm d}^2\bar{\theta}\,
\oint_{\cal C} \frac{{\rm d} \zeta} {2 \pi i \zeta}\;
H(\eta, \zeta)\ ,
\label{eq-action}
\end{equation}
where ${\cal C}$ is a contour in the complex $\zeta$-plane that
generically depends on the singularity structure of the superspace 
density $H$. We use the conventions for the $N=2$
projective superfields $\eta$ as in Appendix B of \cite{DWRV}.

The question is now which function $H$ corresponds to the QK metric
(\ref{QK-metric}). Since for the tree-level c-map, there are (at
least) $n+2$ commuting isometries generated by constant shifts of
$B_I$ and $\sigma$ in (\ref{QK-metric}), we can describe the action in terms
of $N=2$ tensor multiplets,
\begin{equation}\label{eta}
\eta^I = \frac{v^I}{\zeta} + G^I - {\bar v}^I \zeta\ ,
\end{equation}
where $v^I$ project to chiral $N=1$ superfields and $G^I$ to linear
superfields that describe $N=1$ tensor multiplets; each of the latter contains a 
real scalar component field\footnote{We will
also denote the real scalar field of an $N=1$ tensor multiplet $G^I$
by $G^I$. It should be clear from the context what is meant.} 
as well as a component tensor. 
That tensor multiplets appear is
no surprise--as noted above, the c-map 
maps vector multiplets directly into tensor multiplets, 
and the pure $N=2$ supergravity
multiplet is mapped into a double-tensor multiplet that is
dual to the universal hypermultiplet.
The component tensor multiplet Lagrangian that appears after the c-map was 
derived in \cite{FS}. More information on the double-tensor multiplet and 
the general $N=2$ scalar tensor multiplet couplings can be found 
in \cite{TV-DTM}. Superspace effective actions in the context 
of type II string compactifications were also discussed in \cite{BS}.

If we start with the IIB theory, the vector multiplet scalars are
identified with the complex structure moduli, \ie the
periods of the holomorphic three-form
(\ref{Om-periods}). After the c-map, on the type IIA side, the same periods
now define scalars in the tensor multiplet sector. Moreover, the coordinates
$G^I$ are associated to the periods of the RR three-form $C$ of
type IIA. So we have
\begin{equation}\label{C-periods}
v^I = \int_{A_I}\,\Omega\ ,\qquad
G^I = \int_{A_I}\,C\ .
\end{equation}
Similarly, there are the symplectically dual periods. Integrating $C$ over the
dual three-cycles $B^I$ yields new scalars that are associated to
the scalars dual to the tensor appearing in each tensor multiplet.

The constraints from superconformal invariance require scale and
$SU(2)_R$ symmetry. This implies that $H$ is
a function homogeneous of first degree\footnote{Actually, quasihomogeneity 
up to terms of the form $\eta\ln(\eta)$ is sufficient\cite{DWRV}, but such 
terms do not seem to arise in the c-map.}  (in $\eta$) 
and without explicit $\zeta$ dependence \cite{DWRV}. The scaling weights are
such that $\eta$ has weight two, and hence $H(\eta)$ has
weight two as well. The $SU(2)_R$ transformations rotate the three scalars
of each tensor multiplet:
\begin{equation}\label{su2-tensor}
\delta v^I =- i\varepsilon^3 v^I +\varepsilon^- G^I\ ,\qquad
\delta G^I = -2(\varepsilon^- {\bar v}^I + \varepsilon^+ v^I)\ ,
\end{equation}
and leave $G^IG^J+2v^I{\bar v}^J+2{\bar v}^Iv^J$ invariant for all
$I$ and $J$.

As we will see later on, the c-map identifies the
inhomogeneous vector multiplet scalars with the inhomogeneous
coordinates $Z^I\equiv v^I/v^1$, and furthermore the coordinates $A^I$ in
(\ref{QK-metric}) will be identified with the imhomogeneous coordinates
associated to $G^I$. Notice the difference in scaling
weights for the homogenous coordinates: the vector multiplet
scalars $X^I$ in (\ref{K-pot}) have conformal weight 1, whereas
the $v^I$ in (\ref{eta}) have weight two. This is the reason why we use a 
different symbol.

\subsection{Legendre transform and hyperk\"ahler potential}

The projective superspace Lagrangian (\ref{eq-action}) defines the theory
in terms of tensor multiplets. It is well known that in four spacetime
dimensions, tensor multiplets can be dualized into
hypermultiplets. In supersymmetric theories, such
a duality can be performed by doing a Legendre transform on the $N=1$
tensor multiplets \cite{Lindstrom:1983rt,Hitchin:1986ea}. We first introduce
the superspace Lagrangian density
\begin{equation}
{\cal L}(v,\bar v,G)\equiv{\rm Im} \oint_{\cal C} \frac{{\rm d} \zeta}
{2 \pi i \zeta}\; H(\eta, \zeta)\ .
\end{equation}
Integrating this over the $N=2$ superspace measure, one gets the action 
(\ref{eq-action}).

The Legendre transform with respect to $G^I$ is defined by
\begin{equation}\label{legendre-tr}
\chi (v,\bar v,w,\bar w) \equiv {\cal L} (v,\bar v,G)  - (w+\bar w)_I\,G^I\ ,
\qquad w_I+\bar w_I = \frac{\partial {\cal L}}{\partial G^I}\ .
\end{equation}
Observe that the $w_I$ have scaling weight zero, since both ${\cal
L}$ and $G$ have weight two. The object $\chi$ is called the
hyperk\"ahler potential, and serves as the K\"ahler potential for
the HKC. It has scaling weight two and is a function of the
complex coordinates $v^I$ and $w_I$.
The HKC metric only depends on $w$ through the combination $w+\bar w$
and the absence of the imaginary parts of $w$ reflects the
commuting isometries. We can rewrite (\ref{legendre-tr}) as a Legendre 
transform on the function $H$,
\begin{equation}\label{chi-H}
\chi(v,\bar v, w,\bar w)={\rm Im}\,\oint_{\cal C}\,\frac{{\rm d}\zeta}
{2\pi i \zeta}\,
\Bigl[H(\frac{{\bar v}^I}{\zeta}+G^I -v^I\zeta) - G^I\frac{\partial H}
{\partial G^I}\Bigr]\ ,
\end{equation}
with the defining relation for the coordinates $w+\bar w$,
\begin{equation}
(w+\bar w)_I={\rm Im} \oint_{\cal C}\,\frac{{\rm d}\zeta}{2\pi i\zeta}\,
\frac{\partial H}{\partial \eta^I}\ .
\end{equation}

Since $H$ is homogeneous of first degree in $\eta$, it follows that the
hyperk\"ahler potential is also homogeneous of first degree in $v$ and 
$\bar v$ in the sense of (we take $\lambda$ real)
\begin{equation}\label{chi-homog}
\chi(\lambda v,\lambda {\bar v},w,\bar w)=\lambda \chi(v,{\bar v},w,\bar w)\ .
\end{equation}
The $SU(2)_R$ symmetry (\ref{su2-tensor}) after the Legendre transform 
acts on the coordinates $v$ and $w$ as \cite{DWRV}
\begin{equation}
\delta v^I =- i\varepsilon^3 v^I +\varepsilon^- G^I(v,{\bar v},w,{\bar w})\ ,
\qquad \delta w_I=\varepsilon^+ \frac{\partial \cal L}{\partial {\bar v}^I}\ ,
\end{equation}
where $G^I$ has to be understood as the function of the coordinates
$v,{\bar v},w,{\bar w}$ obtained by the Legendre transform defined in
(\ref{legendre-tr}). The coordinates $w_I$ do not transform under
$\varepsilon^3$. One can now explicitly check that the hyperk\"ahler
potential is $SU(2)_R$ invariant,
\begin{equation}
\delta \chi = {\cal L}_{v^I}\,\delta v^I
+{\cal L}_{{\bar v}^I}\,\delta {\bar v}^I
-\delta (w_I+{\bar w}_I)\, G^I=0\ .
\end{equation}
(The $\delta G$ terms cancel identically because $\chi$ is a Legendre transform).
For the generators $\varepsilon^{\pm}$ this is immediately obvious; for
variations proportional to $\varepsilon^{3}$ one needs to use the
invariance of ${\cal L}$, \ie $v^I{\cal L}_{v^I} = {\bar v}^I{\cal L}_{{\bar v}^I}$.

\subsection{Hints from the rigid c-map}

The Quaternion-K\"ahler space in the image of the c-map has
dimension $4(n+1)$. The hyperk\"ahler cone above it, which appears in the
off-shell superspace formulation, has dimension $4(n+2)$.
It therefore needs to be described by $n+2$ tensor
multiplets, say $\eta^I$ and $\eta^0$, where $I=1,\cdots , n+1$.
As we will show, the answer for the tree level c-map is
given by
\begin{equation}\label{G-cmap}
H(\eta^I,\eta^0) = \frac{F (\eta^I)}{\eta^0}\ ,
\end{equation}
where $F$ is the prepotential of the special K\"ahler geometry, now evaluated
on the tensor multiplet superfields $\eta$. This is our main result.
Note that this $H$ does not depend explicitly on $\zeta$ and is homogeneous of degree one, as required by superconformal invariance.

In the next section we give a detailed proof of (\ref{G-cmap}) by explicit
calculation; here we give the several intuitive arguments.
In \cite{CFG,ghk,DWRV}, the rigid c-map, a
map from a $2(n+1)$-dimensional rigid special K\"ahler manifold with arbitrary
(not necesserily homogeneous of second degree) holomorphic
prepotential $F(X^I)$, to a $4(n+1)$-dimensional hyperk\"ahler space is 
discussed.
The c-map can again be formulated in terms of tensor multiplets, and
the resulting projective superspace description is based on the function
\begin{equation}\label{rcmap}
{\rm {rigid\ c-map :}} \qquad H(\eta^I,\zeta) =
\frac{F(\zeta \eta^I)}{\zeta^2}\ .
\end{equation}
The proof was given in \cite{ghk,DWRV} by explicitly doing the contour
integral with the contour around the origin. We give a
more direct derivation in Appendix A, where the rigid c-map is performed
in $N=1$ superspace.

When $F$ is homogeneous of second degree, the $\zeta$-dependence drops out of
(\ref{rcmap}):
\begin{equation}\label{G=F}
H = F(\eta^I)\ .
\end{equation}
The $\zeta$-independence is one of the two requirements of $N=2$ superconformal
symmetry. The other requirement is that $H$ is homogenous of first degree
(in $\eta$), but this is clearly not the case in (\ref{G=F}).
We can fix this problem
by introducing a compensator $\eta^0$ that restores the correct homogeneity.
The resulting hyperk\"ahler space is then $4(n+2)$ dimensional, but this
dimensionality is precisely what is needed for the {\it local} c-map ! So we
are led to consider (\ref{G-cmap}),
which this defines an HKC above a QK manifold. All constraints from $N=2$
superconformal symmetry are satisfied.

For the universal hypermultiplet, the corresponding holomorphic
prepotential is quadratic, and there is only one (compensating) vector
multiplet, hence $F(X^1)=(X^1)^2$. After the c-map, we get a description in
terms of two tensor multiplets:
\begin{equation}
F(\eta^0,\eta^1) = \frac{(\eta^1)^2}{\eta^0}\ .
\end{equation}
We know from explicit calculations in \cite{DWRV} that this is the right
answer.

\section{Hypermultiplet formulation}
\setcounter{equation}{0}

We now explicitly compute the QK geometry corresponding
to the superspace density
\begin{equation}\label{G-F}
H(\eta)=\frac{F(\eta^I)}{\eta^0}\ .
\end{equation}
We show that the result matches exactly with the QK manifolds obtained by
the c-map, \ie we prove that (\ref{G-F}) leads to (\ref{QK-metric}).

\subsection{Gauge fixing and contour integral}

As explained in the previous section, any hyperk\"ahler cone has a local 
$SU(2)_R$ and
dilatation symmetry. In general, these symmetries are nonlinearly realized.
On the tensor multiplet side, these symmetries act linearly:
the $SU(2)_R$ rotates the three scalars inside each tensor multiplet, and
the dilatations uniformly rescale them with weight two. The dilatations and
$U(1)\subset SU(2)$ are
usually combined together, as well as the remaining two generators
(say $T_{\pm}$), into complex generators. To evaluate the contour integral, it
is convenient to first impose some gauge choices. For the $T_{\pm}$
symmetries we choose the gauge
\begin{equation}
v^0 = 0 \ ,
\end{equation}
where $v^0$ is the chiral $N=1$ superfield sitting in the
compensator $\eta^0$. In this gauge, we have that $\eta^0=G^0$
and this simplifies the pole structure in the complex
$\zeta$-plane. Note that this gauge choice does not
impose restrictions on the periods (\ref{C-periods}), since in that formula
the label $I$ does not contain 0. The (complexified) dilatations can
be fixed by choosing $v^1=1$, but we postpone implementing
this gauge.

The contour integral that we now have to evaluate is given by
\begin{equation}
{\cal L} (v,{\bar v},G)= \frac{1}{G^0}\, {\rm Im}\ \oint
\frac{{\rm d} \zeta}{2\pi i} \frac{F(\zeta \eta^I)}{\zeta^3}\ .
\end{equation}
We have
\begin{equation}
\zeta \eta^I={v}^I+\zeta G^I - \zeta^2 {\bar v}^I\ ,\qquad
I=1,\cdots , n+1\ ,
\end{equation}
which, for nonzero values of $v$, has no zeroes at $\zeta=0$.
Therefore, assuming $F$ is regular at $\eta=0$, $F(\zeta \eta)$ 
has no poles (in $\zeta$) inside the contour
around the origin (the same reasoning was used for
the rigid c-map). It is now easy to evaluate the contour integral, because
the residue at $\zeta=0$ replaces all the $\zeta \eta^I$ by $v^I$.
The result is
\begin{equation}\label{cont-F}
{\cal L}(v,\bar v,G)=\frac{1}{4G^0}\Big( N_{IJ}G^I G^J - 2 K \Big)\ ,
\end{equation}
where $K(v,\bar v)$ is the K\"ahler potential of the rigid special
geometry given in (\ref{K-pot}), with $F_I$ now the derivative
with respect to $v^I$ . Because the $v^I$ and $G^I$ have conformal weight
two, $F(v)$ and $K(v,\bar v)$ have weight four, ${\cal L}$ has
weight two and $N_{IJ}$ has weight zero. The function
${\cal L}$ satisfies the Laplace-like equations \cite{PSS1,PSS2,Hitchin:1986ea}
\begin{equation}
{\cal L}_{G^IG^J}+{\cal L}_{v^I{\bar v}^J}=0\ .
\end{equation}
The equation is not satified for the components ${\cal L}_{G^0G^0}$
and ${\cal L}_{G^0G^I}$, because we have chosen a gauge,
$v^0=0$. It would be interesting to compute ${\cal L}$ for arbitrary
values of $v^0$. For the universal hypermultiplet, this was done
in \cite{Anguelova:2004sj}. 

\subsection{The hyperk\"ahler potential}

To compute the hyperk\"ahler potential, we have to dualize the tensor
multiplets into hypermultiplets. As described in the previous section,
in supersymmetric theories, such a duality can be performed
by doing a Legendre transform. Although this can be done in
$N=2$ projective superspace, here we perform the duality
in $N=1$ superspace by dualizing the $N=1$ tensor multiplets
$G^I$ into $N=1$ chiral superfields \cite{Lindstrom:1983rt,Hitchin:1986ea}:
\begin{equation}\label{legendre-chi}
\chi (v,\bar v,w,\bar w) = {\cal L}(v,\bar v,G) +(w+{\bar w})_0\,G^0  -
(w+{\bar w})_I\,G^I \ ,
\end{equation}
The hyperk\"ahler
potential $\chi$, computed by extremizing\footnote{The relative minus signs between the
last two terms in (\ref{legendre-chi}) is purely a matter of
convention.} (\ref{legendre-chi}) with respect  $G^0,G^I$ 
completely determines the hypermultiplet theory and its
associated hyperk\"ahler geometry. In general, it is a function of the $2(n+2)$
complex coordinates $v^0,v^I$ and $w_0,w_I$, but we
have already gauge-fixed $v^0=0$.
The geometry of the HKC only depends on $w$ through the
combination $w+\bar w$ which makes manifest
the $n+2$ commuting isometries. The Legendre transform of (\ref{cont-F})
gives:
\begin{equation}\label{LT-phi}
\frac{G^I}{G^0}=2N^{IJ}(w+{\bar w})_J\ ,\qquad (G^0)^2=
\frac{K}{2\Bigl((w+{\bar w})_IN^{IJ}(w+\bar w)_J-(w+{\bar w})_0\Bigr)}\ .
\end{equation}
Up to an irrelevant overall sign, we find, using (\ref{cont-F})
\begin{equation}\label{chi-phi}
\chi \Bigl(v,\bar v, G(v,\bar v,w,\bar w)\Bigr)=\frac{K(v,\bar
v)}{G^0}\ ,
\end{equation}
where $G^0$ is determined by (\ref{LT-phi}). More explicitly,
in terms of the HKC coordinates,
\begin{equation}\label{cmap-HKC}
\chi (v,\bar v, w,\bar w)= {\sqrt 2}\,\,{\sqrt {K(v,\bar v)}
}\,{\sqrt {(w+{\bar w})_IN^{IJ}(w+\bar w)_J-(w+{\bar w})_0}}\ .
\end{equation}
The reader might be surprised that in the HKC coordinates, the
hyperk\"ahler potential is proportional to the {\it square-root} of
the special K\"ahler potential $K$. This is completely fixed by the
scaling weights:
The last factor on the right hand side has
weight zero; since $K(v,\bar v)$ has weight four (as opposed to
$K(X,\bar X)$ which has weight two), the square-root is needed to
give $\chi$ scaling weight two. Similarly, for (\ref{chi-phi}),
the weights work out correctly because $G^0$ has weight two.

\subsection{Twistor space}
The twistor space above a $4(n+1)$ dimensional QK has dimension
two higher, and is K\"ahler. It can be seen as a CP$^1$ bundle
over the QK. It can also be obtained from the HKC by gauge fixing
dilatations and $U(1)\subset SU(2)$ \cite{DWRV}.
Instead of fixing a gauge, we use inhomogeneous coordinates $Z^I$ as
defined in (\ref{proj-coord}). This allows us to 
choose different gauges. The K\"ahler potential of the rigid special
K\"ahler manifold can be written as
\begin{equation}\label{K-calK}
K=|X^1|^2\, {\rm e}^{\cal K}\ ,
\end{equation}
with ${\cal K}$ given by (\ref{local-K}). The same equation holds
in the variables $v^I$, and we define inhomogeneous coordinates as
\begin{equation}
Z^I=\frac{v^I}{v^1}=\{1,Z^A\}\ ,
\end{equation}
where $A$ runs over $n$ values. As we show below, these
inhomogeneous coordinates will be identified with
(\ref{proj-coord}).

The K\"ahler potential on the twistor space, denoted by $K_T$, is
given by the logarithm of the hyperk\"ahler potential \cite{DWRV}:
\begin{equation}\label{ktzw}
K_T(Z,\bar Z,w,\bar w)=\frac{1}{2}\Big[ {\cal K}(Z,\bar Z) 
+ {\rm ln} \Big((w+{\bar
w})_IN^{IJ}(w+\bar w)_J-(w+{\bar w})_0\Big) \Big] +{\rm ln}({\sqrt
2})\ .
\end{equation}

On the twistor space, there always exists a holomorphic one-form
${\cal X}$ which can be constructed from the holomorphic two-form
that any hyperk\"ahler manifold admits. In our case this one-form
is obtained from the holomorphic HKC two-form $\Omega = {\rm d}w_I
\wedge {\rm d}v^I$. Without going into details, it is given by
\cite{DWRV}
\begin{equation}
{\cal X}=2Z^I{\rm d}w_I\ .
\end{equation}
The metric on the QK manifold can then be computed\footnote{Note that the constant
term in $K_T$ (\ref{ktzw}) enters in (\ref{G-metric}).}:
\begin{equation}\label{G-metric}
G_{\alpha{\bar \beta}}=K_{T,\,\alpha {\bar \beta}}-{\rm
e}^{-2K_T}{\cal X}_\alpha {\bar {\cal X}}_{{\bar \beta}}\ ,
\end{equation}
where the indices $\alpha,\beta=1,\cdots ,2(n+1)$. The coordinates
$z^\alpha$ on the QK consist of $w_I,w_0$ and the
inhomogeneous coordinates $Z^A$ of the special K\"ahler space. In
total this gives $2(n+1) + 2 + 2n = 4(n+1)$--the (real)
dimension of the QK.

\subsection{The quaternionic metric}

We now compute the QK metric that follows from the 
c-map using (\ref{G-metric}). To compare with (\ref{QK-metric})
we only need to identify the coordinates $w_I,w_0$ with those of
(\ref{QK-metric}), since the $Z^A$ coordinates of the special
K\"ahler manifold can be identified with the ones above. We define
\begin{eqnarray}\label{QK-coord1}
w_0&=&iA^IA^JF_{IJ}-i(\sigma+\frac12A^IB_I)-{\rm e}^\phi\ ,\nonumber\\
w_I&=&iF_{IJ}A^J-\frac{i}{2}B_I\ .
\end{eqnarray}
The metric can be written in these coordinates, and after calculating
we obtain the following result:
\begin{eqnarray}\label{QK-metric2}
{\rm d}s^2&=&{\rm d}\phi^2 - {\rm e}^{-\phi}({\cal N}+{\bar {\cal
N}})^{-1\,IJ}
\Big|2{\cal N}_{IK}{\rm d}A^K+i{\rm d}B_I\Big|^2\nonumber\\
&& + {\rm e}^{-2\phi} \Big({\rm d}\sigma - \frac12(A^I {\rm d}B_I-B_IdA^I)\Big)^2
-4 {\cal K}_{A\bar B}\,{\rm d}Z^A {\rm d}{\bar Z}^{\bar B}\ .
\end{eqnarray}
We have left out an overall normalization constant ($(-1/8)$ to be
precise), and the matrix ${\cal N}_{IJ}$ is defined as in
(\ref{curlyN}). How this matrix comes out of our calculation is
somewhat nontrivial, and we have used the identity (see appendix B
of \cite{DWVVP})
\begin{equation}
({\cal N}+{\bar {\cal N}})^{-1\,IJ}=N^{IJ}-\frac{X^I{\bar
X}^J+{\bar X}^IX^J} {(XN\bar X)}\ .
\end{equation}
The two terms on the right hand side follow
from the two terms on the right hand side of (\ref{G-metric}).
It is then easy to see that this metric coincides with
(\ref{QK-metric}). This concludes the
proof of (\ref{G-F}).

Finally, from (\ref{LT-phi}) we can easily derive the following relations
between the QK coordinates and the tensor multiplet coordinates:
\begin{equation}\label{expr-dilaton}
2A^I =\frac{G^I}{G^0}\ ,\qquad 4{\rm e}^\phi = \frac{K}{(G^0)^2}\ .
\end{equation}
This associates directly the coordinates $A^I$ with the periods
as defined in (\ref{C-periods}). This is consistent with the supergravity
analysis in \cite{DWVVP} where it was shown that the scalars $(A^I, B_I)$
form a symplectic pair, whereas the dilaton is an invariant.

\section{Relation to black holes and topological strings}\label{topstring}
\setcounter{equation}{0}
\subsection{Recap}

What we have established in the previous sections is that the
tensor multiplet Lagrangian at tree-level (in the string coupling
constant $g_s$) is determined by the prepotential
$F(X^I)$ of special geometry via the c-map. The action can be written in projective 
superspace as
\begin{equation}\label{the-action}
S_{{\rm tensor}}= {\rm Im} \int\,{\rm d}^4 x \; {\rm d}^2\theta 
{\rm d}^2\bar{\theta}\,\oint_{\cal C} \frac{{\rm d} \zeta} {2 \pi i \zeta}\; 
\frac{F(\eta^I)}{\eta^0}\ ,
\end{equation}
where $\eta(\zeta)$ are $N=2$ tensor multiplets defined in (\ref{eta}),
consisting of an $N=1$ chiral multiplet $v$ and an $N=1$ tensor multiplet
$G$. The action (\ref{the-action}) has an $SU(2)_R$ symmetry, and in the 
gauge $v^0=0$ the contour integral can be done most easily:
\begin{equation}\label{tensor-action}
S_{{\rm tensor}}= \int\,{\rm d}^4 x \; {\rm d}^2\theta 
{\rm d}^2\bar{\theta}\,\,\,\frac{1}{4G^0}\Big(N_{IJ}G^IG^J-2K(v,\bar v)\Big)\ .
\end{equation}
To get the tree-level hypermultiplet
action, we need to Legendre transform (\ref{tensor-action}) with respect 
to the $N=1$ tensor multiplets $G^I$ and $G^0$. The resulting function
was given in (\ref{cmap-HKC}) and is called the hyperk\"ahler potential 
$\chi$. It can be written in a more compact form as
\begin{equation}\label{chi-G0}
\chi \Bigl(v,\bar v, G(v,\bar v,w,\bar w)\Bigr)=\frac{K(v,\bar
v)}{G^0}\ ,
\end{equation}
where $G^0(v,\bar v,w,\bar w)$ is determined by (\ref{LT-phi}). 
One can now make use of the homogeneity properties of the hyperk\"ahler 
potential (\ref{chi-homog}). Introducing the {\it weight one} coordinates
\begin{equation}\label{new-X}
X^I(v,\bar v,w,\bar w)\equiv \frac{v^I}{{\sqrt {G^0(v,\bar v,w,\bar w)}}}\ ,
\end{equation}
we can conveniently rewrite the hyperk\"ahler potential as
\begin{equation}\label{chi-is-K}
\framebox{$\chi(v,\bar v,w,\bar w)=K\Big(X^I(v,\bar v,w,\bar w),{\bar X}^I(v,\bar v,w,\bar w)\Big)$}\ .
\end{equation}
Here $K$ is the K\"ahler potential of the rigid special geometry, 
$K(X,\bar X)=i({\bar X}^IF_I-X^I{\bar F}_I)$ and the $X^I$ scale with weight
one. There is thus a very simple rule to obtain the HKC hyperk\"ahler potential
from the special K\"ahler potential: just replace the holomorphic coordinates
$X^I$ by the functions $X^I(v,\bar v,w,\bar w)$ as defined by 
(\ref{new-X})!

The relations (\ref{new-X}) and (\ref{chi-is-K}) are written in 
an $N=1$ or component language. The corresponding relations in $N=2$ language 
can best be formulated in terms of tensor multiplets, and are simply 
the statement that the function $H$ is related to 
the prepotential $F$ as in (\ref{G-F}). Indeed, as in (\ref{new-X}), 
if we define {\it weight one} $N=2$ multiplets
\begin{equation}
X^I(\eta) \equiv \frac{\eta^I}{{\sqrt {\eta^0}}}\ ,
\end{equation}
we can write the tensor multiplet superspace density $H$ as
\begin{equation}\label{H-is-F}
\framebox{$H(\eta^I,\eta^0)=F\big[X^I(\eta^I,\eta^0)\big]$}\ .
\end{equation}
Notice that both formulas (\ref{chi-is-K}) and
(\ref{H-is-F}) are consistent with the scaling weights and homogeneity 
properties. There is still a 
(complexified) dilatation gauge that we could choose, for instance
$G^0=1$, or $v^1=1$, but we prefer not do so to keep our formulas are valid
in any gauge. Furthermore, note that the hyperk\"ahler potential 
$\chi$ is a scalar function under symplectic transformations induced by the 
vector multiplet theory. Under symplectic transformations, $(X^I,F_I)$
transforms linearly under the symplectic group, such that $K$ is a
scalar. On the other hand, $H$ is not a scalar, but transforms like the 
prepotential $F$.

\subsection{Legendre transform and black holes}

We now repeat some aspects of the Legendre transform as done in
\cite{Ooguri:2004zv}, see also
\cite{Verlinde:2004ck,OVV}. Consider the identity
\begin{equation}
-\frac{1}{2}|\lambda|^2K(X,\bar X)= {\rm
Im}\Big(\lambda^2X^IF_I\Big) -2\, {\rm Im}(\lambda X^I)\, {\rm
Re}(\lambda F_I)\ ,
\end{equation}
for any complex quantity $\lambda$. Using the homogeneity property
of $F$, we can rewrite this identity as
\begin{equation}
-\frac{1}{4}|\lambda|^2K(X,\bar X)= {\rm Im}[\lambda^2F( X^I)] -\,
{\rm Im}(\lambda X^I)\, {\rm Re}(\lambda F_I( X^I))\ .
\end{equation}
We now define real quantities $p^I,\phi^I,q^I$ and ${\cal F}$ by
\begin{equation}\label{def-pq}
p^I+\frac{i}{\pi}\phi^I\equiv \lambda X^I\ , ~~~ q_I\equiv {\rm
Re}(\lambda F_I( X^I))\ , ~~~ {\cal F}\equiv-\pi{\rm Im}[\lambda^2F( X^I)]
= -\pi{\rm Im}[F(\lambda X^I)]\ .
\end{equation}
They satisfy the relation
\begin{equation}
q_I=-\frac{\partial  {\cal F}}{\partial \phi^I}\ ,
\end{equation}
which can be used to write the $\phi^I$ as a function of $q_I$ and $p^I$. 
Now it follows that the K\"ahler potential is the Legendre
transform of ${\cal F}$:
\begin{equation}\label{K-F}
\frac{\pi}{4}K(p,q)={\cal F}[p,\phi(p,q)]+\phi^I(p,q)\,q_I\ .
\end{equation}
Notice that we have not used BPS-like equations whatsoever to
derive this relation. In the context of black holes, the $q_I$ and $p^I$ 
are of course related to the electric and magnetic charges of the black hole
via the attractor equations.

Using (\ref{chi-is-K}), we can now write the hyperk\"ahler potential
as a Legendre transform,
\begin{equation}\label{chi-is-LT}
\frac{\pi}{4} \,\chi (p,q) ={\cal F}[p,\phi(p,q)] +\phi^I(p,q)q_I\ .
\end{equation}
The only thing we have to do in this equation is to interpret the ``charges''
$p^I$ and $q_I$ in terms of the hypermultiplet variables, \ie 
\begin{equation}
p^I+\frac{i}{\pi}\phi^I=\frac{\lambda v^I}{\sqrt {G^0(v,\bar v,w,\bar w)}}\ ,
\end{equation}
and similarly for $q_I$.

Could this have any relation with the Legendre transform as defined in
(\ref{legendre-tr})? As they stand, the two Legendre transforms
seem totally different, since in (\ref{legendre-tr}) one Legendre
transforms ${\cal L}$ with respect to the tensor multiplet variables $G^I$,
whereas in (\ref{chi-is-LT}) one Legendre transforms ${\cal F}$ with respect
to the ${\rm Im}(X^I)$. 
On the other hand, the relations (\ref{chi-is-K})
and (\ref{H-is-F}) do connect them. Furthermore,
the $SU(2)_R$ transformations rotate $G^I$ into ${\rm Im}( v^I)$; since we work
in the gauge $v^0=0$, this symmetry is not manifest, but it should
allow us to rotate the two Legendre transforms into each other. 

We now give an additional argument.
As discussed in \cite{OVV} (which involves an interpretation of the OSV
wave function in the black hole context \cite{Ooguri:2004zv}) 
the Hartle-Hawking wave function for black holes is a function, not only of 
the moduli of the Calabi-Yau, but also 
of the Ramond-Ramond gauge potentials, which in this paper we have denoted 
by $G^I$.  
However, the notion of the mini-superspace used in \cite{OVV} 
amounts to choosing a reduction to the BPS sector
of the theory. In such a case $G^I$ (to be more precise, $G^I/{\sqrt {G^0}}$) 
gets identified with ${\rm Im}X^I$.  
So in principle we can view the topological string wave function as a 
function of ${\rm Re} X^I+i G^I$. Using the relations between the topological 
string amplitude and the prepotential $\lambda X^1=\frac{4\pi i}
{g_{{\rm top}}}$ and ${\cal F}=F_{{\rm top}}+{\bar F}_{{\rm top}}$, we can
write
\begin{equation}
\psi_{{\rm {black\ hole}}}= 
{\rm e}^{F_{{\rm top}}\big[{\rm Re}X^I+iG^I\big]}\ .
\end{equation}
In this case in order to obtain the black hole entropy we have to consider the
Fourier transform (which to leading order is the Legendre transform) of 
$|\psi|^2$ with respect to $G^I$.  This then is exactly of the same 
structure as in (\ref{legendre-tr}) or (\ref{legendre-chi}). The question is then 
why should a formula resembling the formula corresponding to entropy of 
black holes be related to our discussion here.

Here we offer a possible explanation, which may be the basis of this 
connection. As discussed in \cite{OVV}, the relation between the 
Hartle-Hawking wave function and topological
strings goes via compactification of the four dimensional theory on a circle
and writing the reduced wave function on possible degrees of freedom, subject
to preserving half the supersymmetry.  However, once we compactify on a circle
the black hole states, running in the extra circle, viewed as Euclidean 
time, play the role of instantons of the three dimensional theory, which by 
T-duality on the radius of the circle gets related to hypermultiplets in 
the 4-dimensional theory, as discussed in the context of c-map.  
Thus the exchange of the role between black hole states and instantons 
may be a partial explanation of this fact. 

\section{Higher derivative terms}
\setcounter{equation}{0}

A natural question is to ask what the higher genus
partition function of the topological string computes for the
tensor or hypermultiplet. This question was addressed in
\cite{Antoniadis:1993ze}, where higher derivative corrections on
the universal hypermulitplet were found that multiply the genus
$g$ partition function. Here we will write down such terms in
superspace by using a similar procedure as for the vector
multiplet action.

The topological A-model computes F-terms in the four-dimensional
supergravity effective action, proportional to
higher powers of the Riemann curvature and graviphoton field strength
\cite{Antoniadis:1993ze,Bershadsky:1993cx}. They can be nicely
encoded using superspace techniques, by putting the vector multiplets
in a chiral background \cite{deWit:1996ix}. In $N=2$ chiral superspace,
vector multiplet actions can be written as
\begin{equation}\label{VM-action}
S={\rm {Im}} \Big( \int {\rm d}^4x\, {\rm d}^4{\theta} F(X) \Big)\ ,
\end{equation}
where $F$ is a holomorphic function, homogeneous of degree two in
$N=2$ restricted chiral superfields $X^I$ (\ie the $N=2$ vector multiplet
superfields). To generate the higher curvature terms, one considers generalized
actions of the type (\ref{VM-action}) by including a background $N=2$
(unrestricted) chiral superfield $\Phi$, which is associated with the square
of the Weyl multiplet ${\cal W}^2$ (having scaling weight 2).
The action is based on a new prepotential $F(X,{\cal W}^2)$ and we
expand it in a power series 
\begin{equation}
F(X^I,{\cal W}^2)=\sum_{g=0}^{\infty} F_g(X^I) ({\cal W}^2)^g\ .
\end{equation}
It then turns out that the coefficient functions $F_g(X)$ are related
to the genus $g$ topological partition function.

We can now set up a similar construction on the tensor multiplet side, 
after having done the c-map. The
Weyl multiplet becomes the universal hypermultiplet after the c-map,
so one indeed expects the higher genus terms to correspond to higher
derivative terms in the universal hypermultiplet only \cite{Antoniadis:1993ze}.
We can immitate the same trick as for the vector multiplets, by putting the
Lagrangian corresponding to (\ref{G-F}) in an (unrestricted) projective
superfield background $\Upsilon$. Based on the c-map, and arguments
given above, we put the prepotential in this background and expand
\begin{equation}
F(\eta^I,\Upsilon)=\sum_{g=0}^{\infty} F_g(\eta^I) (\Upsilon)^g\ .
\end{equation}
A similar expansion also appears in \cite{BS}.
The background $\Upsilon$ should be identified with the c-map of the
(square of the) Weyl multiplet ${\cal W}^2$, and is describing the universal
hypermultiplet. The coefficient functions $F_g$ are again be related
to the genus $g$ partition function of the topological string.
We take
\begin{equation}
\Upsilon = \nabla^2 {\bar \nabla}^2 \, (L^2)^{1/2}\ ,
\end{equation}
where $L^2$ is an appropriate function of the tensor 
multiplet describing the universal hypermultiplet. The $\nabla$ and 
$\bar \nabla $ operators are there to generate the higher derivatives, in such
a way that (powers of) $\Upsilon$ can be integrated over superspace, and such 
that the the scaling and $SU(2)_R$ symmetries are preserved. 
A candidate would be
\begin{equation}
L^2 = L^I_{ij}N_{IJ}L^{J\,ij}\ ,
\end{equation} 
where the $L_{ij}$ describe the components of a tensor multiplet $\eta$ by 
means of $L_{+-}\propto G, L_{++} \propto v$. That this multiplet 
contains the dilaton can be argued from (\ref{expr-dilaton}). 
It remains to be shown that this leads to the same answer as in 
\cite{Antoniadis:1993ze}, where higher derivative terms 
for the hypermultiplets were written down in components. 
A more systematic treatment of higher derivative terms for tensor multiplets
in components will be given in \cite{DWSV}.
\vskip .5cm
\noindent{\bf Acknowledgements}

\noindent It is a pleasure to thank Bernard de Wit, Boris Pioline and 
Frank Saueressig for stimulating discussions. 
Most of this work has been initiated and completed during the 
2004 and 2005 Simons Workshops in Physics and Mathematics. 
SV and CV thank the C.N. Yang Institute for Theoretical Physics
and the Department of Mathematics at Stony Brook University for hosting the
workshops and for partial support. MR is supported in part by
NSF grant no.~PHY-0354776. CV is supported
in part by NSF grants PHY-0244821 and DMS-0244464. 

\appendix
\section{The rigid c-map}
The classical rigid c-map can be easily understood by reducing $N=2$ $D=4$
superspace to $N=1$ superspace
and then going down to $D=3$; the relation is also direct and
transparent in $N=2$ superspace, but the projective formalism is less familiar \cite{ghk,DWRV}.

Consider an $N=2$ vector-multiplet Lagrange density:
\be
\Lag_{vect}=-{\rm Im}\left[\int {\rm d}^4\q F(X^I)\right]~,
\ee
where the measure ${\rm d}^4\q$ is the $N=2$ chiral measure. In $N=1$ superspace, this becomes
\be\label{n1v}
\Lag_{vect}=-{\rm Im}\left[\int {\rm d}^2\q \left(\frac{\pa F}{\pa X^I}
\bar D^2\bar X^I+
\frac{\pa^2 F}{\pa X^I\pa X^J}W^{I\al}W^J_\al\right)\right]~,
\ee
where $X^I$ are now $N=1$ chiral superfields and $W^{I\al}$ is the $N=1$ vector multiplet
field strength. Descending to to $D=3$ does not change (\ref{n1v}) {\em except} that $W^{I\al}$
can now be written in terms of a real lower dimension field strength $G^I$:
\be\label{wg}
W^I_\al=\frac{i}{\sqrt{2}}\bar D_\al G^I~~~,~~ D^2G^I=\bar D^2G^I=0~;
\ee
this is not possible in $D=4$ because it violates Lorentz invariance--$W_\al$ and $\bar D_{\dot\al}$
transform in conjugate representations of $Sl(2,\mathbb{C})$ 
that reduce to the same representation of $Sl(2,\mathbb{R})$.
Thus we find the $D=3$ Lagrange density:
\be
\Lag_{vect}={\rm Im}\left[\int {\rm d}^2\q \left(-\frac{\pa F}{\pa X^I}\bar D^2\bar X^I
+\frac12\frac{\pa^2 F}{\pa X^I\pa X^J}\bar D^\al G^I\bar D_\al G^J_\al\right)\right]~.
\ee
Using the chirality constraints on $X^i$ and the linear constraints (\ref{wg}) on $G$, we can rewrite
this as a full superspace integral:
\be\label{lv}
\Lag_{vect}=\int {\rm d}^2\q\, {\rm d}^2\bar\q~ {\rm Im}\left(-\frac{\pa F}
{\pa X^I}\bar X^I+\frac12
\frac{\pa^2 F}{\pa X^I\pa X^J}G^IG^J\right)~.
\ee

We now compare this to the hypermultiplet action. We use the tensor multiplet projective superspace
description of the $D=4$ hypermultiplet; this involves superfields $\eta^I=\frac{X^I}\zeta+G^I-\zeta\bar X^I$ which are real under the composite operation of complex conjugation and the antipodal map
$\bar\zeta\to\frac{-1}\zeta$ on $\mathbb{CP}^1$. The general tensor multiplet
$D=4$ Lagrange density is
\be\label{lh}
\Lag_{hyper}=\int {\rm d}^2\q\, {\rm d}^2\bar\q\oint\frac{{\rm d}\zeta}
{2\pi i\zeta} G(\eta^I,\zeta)
~~~,~~\mathcal{R}(G)=G~.
\ee
Now consider the special case when
\be
G={\rm Im}_\mathcal{R}\left(\frac{F(\zeta\eta^I)}{\zeta^2}\right)
\equiv -i\left[\frac{F(\zeta\eta^I)}{\zeta^2}-\zeta^2{\bar F}
(\frac{-\eta^I}\zeta)\right]~,
\ee
where ${\rm Im}_\mathcal{R}$ means the imaginary part with respect to the composite conjugation $\mathcal{R}$. Consider the first term
\be
\frac{F(\zeta\eta^I)}{\zeta^2}\equiv\frac{F(X^I+\zeta G^I-\zeta^2\bar X^I)}
{\zeta^2}~;
\ee
for $F(X^I)$ regular at $X^I=0$, the contour integral gets just two contributions:
\be
-\frac{\pa F}{\pa X^I}\bar X^I+
\frac12\frac{\pa^2 F}{\pa X^I\pa X^J}G^IG^J~.
\ee
Plugging this into (\ref{lh}) gives precisely $\Lag_{vect}$ (\ref{lv}).

\raggedright


\begin{thebibliography}{99} 

\bibitem{Bershadsky:1993cx}
M.~Bershadsky, S.~Cecotti, H.~Ooguri and C.~Vafa, {\it Kodaira-Spencer
theory of gravity and exact results for quantum string amplitudes},
Commun.\ Math.\ Phys.\  {\bf 165} (1994) 311, \texttt{hep-th/9309140}.

\bibitem{Antoniadis:1993ze}
I.~Antoniadis, E.~Gava, K.~S.~Narain and T.~R.~Taylor, {\it Topological
amplitudes in string theory}, Nucl.\ Phys.\ B {\bf 413} (1994) 162,
\texttt{hep-th/9307158}.

\bibitem{Ooguri:2004zv}
H.~Ooguri, A.~Strominger and C.~Vafa, {\it Black hole attractors and the
topological string}, Phys.\ Rev.\ D {\bf 70} (2004) 106007,
\texttt{hep-th/0405146}.

\bibitem{LopesCardoso:1998wt}
G.~Lopes Cardoso, B.~de Wit and T.~Mohaupt, {\it Corrections to macroscopic 
supersymmetric black-hole entropy}, Phys.\ Lett.\ B {\bf 451} (1999) 309,
\texttt{hep-th/9812082};  
{\it Deviations from the area law for supersymmetric black holes}, 
Fortsch.\ Phys.\  {\bf 48} 
(2000) 49, \texttt{hep-th/9904005}; 
{\it Macroscopic entropy formulae and non-holomorphic corrections for 
supersymmetric black holes}, Nucl.\ Phys.\ B {\bf 567} (2000) 87,
\texttt{hep-th/9906094}.

\bibitem{BBS}
K.~Becker, M.~Becker and A.~Strominger,
{\it Five-branes, membranes and nonperturbative string theory}, 
Nucl.\ Phys.\ B {\bf 456} (1995) 130, {\texttt hep-th/9507158}.

\bibitem{Antoniadis:2003sw}
I.~Antoniadis, R.~Minasian, S.~Theisen and P.~Vanhove, {\it String loop 
corrections to the universal hypermultiplet}, Class.\ Quant.\ Grav.\  {\bf 20} 
(2003) 5079, \texttt{hep-th/0307268};

\bibitem{Anguelova:2004sj}
L.~Anguelova, M.~Ro\v{c}ek and S.~Vandoren, {\it Quantum corrections to the 
universal hypermultiplet and superspace}, Phys.\ Rev.\ D {\bf 70} (2004) 
066001, \texttt{hep-th/0402132}.

\bibitem{Davidse:2004gg}
M.~Davidse, U.~Theis and S.~Vandoren, {\it Fivebrane instanton corrections 
to the universal hypermultiplet}, Nucl.\ Phys.\ B {\bf 697} (2004) 48,
\texttt{hep-th/0404147}.

\bibitem{Davidse:2005ef}
M.~Davidse, F.~Saueressig, U.~Theis and S.~Vandoren, {\it Membrane instantons 
and de Sitter vacua}, JHEP {\bf 0509} (2005) 065, \texttt{hep-th/0506097}.

\bibitem{Greene:1996dh}
B.~R.~Greene, D.~R.~Morrison and C.~Vafa, {\it A geometric realization of 
confinement}, Nucl.\ Phys.\ B {\bf 481} (1996) 513, \texttt{hep-th/9608039}.

\bibitem{Ooguri:1996me}
H.~Ooguri and C.~Vafa, {\it Summing up D-instantons}, Phys.\ Rev.\ Lett.\  
{\bf 77} (1996) 3296, \texttt{hep-th/9608079}.

\bibitem{CFG} 
S.~Cecotti, S.~Ferrara and L.~Girardello, {\it Geometry of
type II superstrings and the moduli of superconformal field theories},
Int.\ J.\ Mod.\ Phys.\ {\bf A4} (1989) 2475.

\bibitem{FS}
S.~Ferrara and S.~Sabharwal, {\it Quaternionic manifolds for type II
superstring vacua of Calabi-Yau spaces}, Nucl.\ Phys.\ {\bf B332} (1990) 317.

\bibitem{DWKV}
B.~de Wit, B.~ Kleijn, and S.~Vandoren, {\it Superconformal hypermultiplets},
Nucl. Phys. {\bf B568} (2000) 475, {\texttt hep-th/9909228}.

\bibitem{DWRV} 
B.~de Wit, M.~Ro\v{c}ek and S.~Vandoren, {\it Hypermultiplets,
hyperk\"ahler cones and quaternion-K\"ahler geometry},
JHEP 0102 (2001) 039, {\texttt hep-th 0101161}.

\bibitem{PSS1}
S.~J~Gates, Jr., C.~Hull and M.~Ro\v{c}ek, {\it Twisted multiplets and
new supersymmetric nonlinear sigma models}, Nucl.\ Phys.\ {\textbf B248}
(1984) 157.

\bibitem{PSS2}
A.~Karlhede, U.~Lindstr\"om and M.~Ro\v{c}ek, {\it Selfinteracting tensor
multiplets in $N=2$ superspace}, Phys.\ Lett.\ {\textbf B147} (1984) 297.

\bibitem{BS} N.~Berkovits and W.~Siegel, {\it Superspace effective actions
for 4D compactifications of heterotic and type II superstrings}, Nucl.\ 
Phys.\ {\bf B462} (1996) 213, {\tt hep-th/9510106}.

\bibitem{Berko}
N.~Berkovits, {\it Conformal compensators and manifest type IIB S-duality},
Phys.\ Lett.\ {\bf B423} (1998) 265, {\texttt hep-th/9801009}.

\bibitem{RSV}
D.~Robles Llana, F.~Saueressig and S.~Vandoren, {\it String loop corrected
hypermultiplet moduli spaces}, to appear.

\bibitem{OVV}
H.~Ooguri, C.~Vafa and E.~P.~Verlinde, {\it Hartle-Hawking wave-function
for flux compactifications}, \texttt{hep-th/0502211}.

\bibitem{Pioline}
M.~Gunaydin, A.~Neitzke, B.~Pioline and A.~Waldron, {\it BPS black holes,
quantum attractor flows and automorphic forms}, \texttt{hep-th/0512296}. 

\bibitem{deWit:1984pk}
B.~de Wit and A.~Van Proeyen, {\it Potentials And Symmetries Of General 
Gauged N=2 Supergravity - Yang-Mills Models}, Nucl.\ Phys.\ B {\bf 245} 
(1984) 89.

\bibitem{Bagger:1983tt}
J.~Bagger and E.~Witten, {\it Matter Couplings In N=2 Supergravity},
Nucl.\ Phys.\ B {\bf 222} (1983) 1.

\bibitem{DWVVP} B. de Wit and A. Van Proeyen, {\it Symmetries of
dual quaternionic manifolds}, Phys. Lett. {\bf B252} (1990) 221.\\
B. de Wit, F. Vanderseypen and A. Van Proeyen, {\it Symmetry
structure of special geometries}, Nucl. Phys. {\bf B400} (1993) 463,
{\tt hep-th/9210068}.

\bibitem{CKVPDFDWG}
E.~ Cremmer, C.~Kounnas, A.~Van Proeyen, J.P.~Derendinger, S.~Ferrara,
B.~de Wit and L.~Girardello, {\it Vector multiplets coupled to $N=2$
supergravity: superhiggs effect, flat potentials and geometric structure},
Nucl.\ Phys.\ {\bf B250} (1985) 385.

\bibitem{Swann}
A.~Swann, {\it Hyperk\"ahler and quaternionic K\"ahler geometry}, Math. Ann.
{\bf 289} (1991) 421.

\bibitem{TV-DTM}
U.~Theis and S.~Vandoren,  {\it Instantons in the double-tensor multiplet},
JHEP {\bf 0209} (2002) 059, \texttt{hep-th/0208145};
{\it N = 2 supersymmetric scalar-tensor couplings},
JHEP {\bf 0304} (2003) 042, \texttt{hep-th/0303048}.

\bibitem{Lindstrom:1983rt}
U.~Lindstrom and M.~Ro\v{c}ek, {\it Scalar tensor duality and N=1, N=2
nonlinear sigma models}, Nucl.\ Phys.\ B {\bf 222} (1983) 285.

\bibitem{Hitchin:1986ea}
N.~J.~Hitchin, A.~Karlhede, U.~Lindstrom and M.~Ro\v{c}ek,  {\it Hyperkahler
metrics and supersymmetry}, Commun.\ Math.\ Phys.\  {\bf 108} (1987) 535.

\bibitem{ghk}
  S.~J.~J.~Gates, T.~Hubsch and S.~M.~Kuzenko,
{\it CNM models, holomorphic functions and projective superspace C-maps},
  Nucl.\ Phys.\ {\bf B557} (1999) 443, {\texttt hep-th/9902211}.

\bibitem{Verlinde:2004ck}
E.~P.~Verlinde, {\it Attractors and the holomorphic anomaly},
\texttt{hep-th/0412139}.

\bibitem{deWit:1996ix}
B.~de Wit, {\it N = 2 electric-magnetic duality in a chiral background},
Nucl.\ Phys.\ Proc.\ Suppl.\  {\bf 49} (1996) 191, \texttt{hep-th/9602060}.

\bibitem{DWSV}
B.~de Wit, F.~Saueressig and S.~ Vandoren, in preparation.

\end{thebibliography}
\end{document}